\documentclass[showpacs,prl,preprintnumbers,amsmath,amssymb,aps,floatfix,twocolumn]{revtex4}
\usepackage{graphicx}
\usepackage{dcolumn}
\usepackage[sort&compress]{natbib}
\usepackage{bm}
\usepackage[dvips]{epsfig}

\begin{document}


\title{Glassy state of native collagen fibril}
\author{S.G. Gevorkian$^{1)}$, A.E. Allahverdyan$^{1)}$, D.S. Gevorgyan$^{2)}$, Chin-Kun Hu$^{3)}$}

\address{$^{1)}$Yerevan Physics Institute, Alikhanian Brothers St. 2, Yerevan 375036, Armenia.\\
$^{2)}$Yerevan State Medical University, Koryun St. 2, Yerevan, 375025, Armenia.\\
$^{3)}$Institute of Physics, Academia Sinica, Nankang, Taipei 11529, Taiwan.}

\begin{abstract}

Our micromechanical experiments show that at physiological temperatures
type I collagen fibril has several basic features of the glassy state.
The transition out of this state [softening transition] essentially
depends on the speed of heating $v$, e.g., for $v=1$ C/min it occurs
around 70 C and is displayed by a peak of the internal friction and
decreasing Young's modulus. The softening transition decreases by 45 C
upon decreasing the heating speed to $v=0.1$ C/min. For temperatures
20-30 C the native collagen fibril demonstrates features of mechanical
glassines at oscillation frequencies 0.1-3 kHz; in particular, the internal friction
has a sharp maximum as a function of the frequency.
This is the first example of biopolymer glassines at physiological temperatures,
because well-known glassy features of DNA and globular proteins are
seen only for much lower temperatures (around 200 K). 

\end{abstract}


  

\maketitle

An important aspect of the globular protein physics is the low-temperature glass
transition experimentally observed at $\approx 200$ K
\cite{morozov_gevorkian,nmr,moessbauer,calor,pressure,x,review_1,review_2}.
The understanding of this transition is achieved by means of combining the
information obtained via different methods: micro-mechanical
experiments \cite{morozov_gevorkian}, NMR \cite{nmr}, Moessbauer
spectroscopy \cite{moessbauer}, calorimetric studies \cite{calor},
pressure release experiments \cite{pressure}, and X-ray scattering of
synchrotron radiation \cite{x}; see \cite{review_1,review_2} for
reviews. Although there still open questions here|concerning, in
particular, the specific role of the hydrated water|the rough physical
picture of the glass transition in globular proteins is constructed 
by an analogy with glass-forming liquids \cite{review_1,review_2} and
synthetic polymers \cite{ferry,frenkel}. In particular, it is believed
that the large-scale conformational motion of proteins freezes at
$\approx 200$ K, analogously to freezing of cooperative motion
in glass-forming liquids \cite{review_1,review_2} and segmental motion
in synthetic polymers \cite{ferry,frenkel}. 

Thus, the glassy features as such are not important in the native state
of globular proteins at physiological temperatures, though the glass
transition at much lower temperatures allows to gain
some understanding of relevant motions in proteins
\cite{review_1,review_2}. 

Here we shall demonstrate via micro-mechanical methods that the native
type I collagen fibril (made of fibrous protein, type I collagen
triple-helices) is in a glassy state at physiological temperatures. This
state is displayed via frequency-dependent visco-elastic characteristics
(the Young's modulus and the damping decrement) of the native fibril.
Upon heating the fibril goes out of the glassy state, a phenomenon known
as the softening transition \cite{ferry,frenkel}. The temperature of
this transition depends essentially on the speed of heating, e.g., for
the standard heating speed $v=1$ C/min it occurs around $\approx 70$ C.
Note that around 70 C the fibril starts to undergo the denaturation
process \cite{haly,miles,tiktopulo,tsereteli}. This process has been
studied by different methods, and some of those methods seems to
indicate that this process is inherently irreversible \cite{miles}. We
saw that upon decreasing the heating speed to $v=0.1$ C/min, the
softening transition takes place at $\approx 25$ C, showing that the
glassy features pertain to the native state of the collagen fibril and
do not directly relate to its denaturation. We shall also confirm 
that the glassy features are not seen for the heat-denaturated
collagen fibril, and contrast features of the native collagen fibril 
to those of globular proteins.

\begin{figure}
\includegraphics[width=8.5cm]{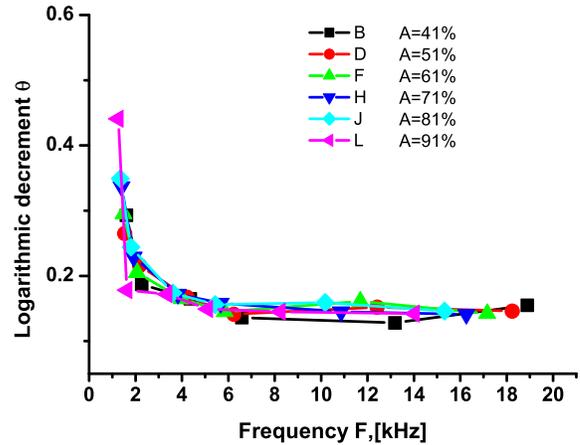}
\caption{ The logarithmic damping decrement $\theta$ of the collagen fibril
versus the frequency at temperature $25^{\circ}$C and varying humidity. }
\label{f1}
\end{figure}

\begin{figure}
\includegraphics[width=8.5cm]{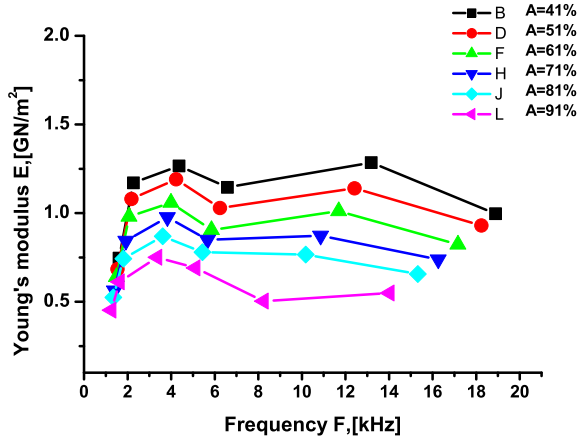}
\caption{  The Young's modulus $E$ of the collagen fibril
versus the frequency at temperature $25^{\circ}$C and varying humidity. }
\label{f2}
\end{figure}

\begin{figure}
\includegraphics[width=8.5cm]{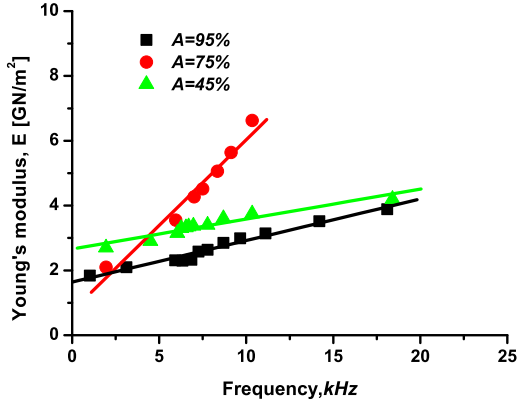}
\caption{The Young modulus $E$ versus frequency of lysozyme crystal for $T=25$ C and varying humidity $A$.}
\label{f5}
\end{figure}

\begin{figure}
\includegraphics[width=8.5cm]{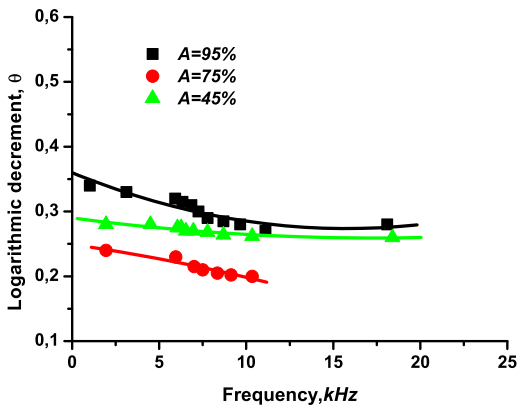}
\caption{The logarithmic damping decrement of lysozyme crystal 
versus frequency for $T=25$ C and varying humidity $A$.}
\label{f6}
\end{figure}

Type I collagen is the major structural element in the extra-cellular
matrix. It forms the basis of fibrous connective tissues, such as
tendon, chord, skin, bones, cornea and dentine; see \cite{fratzl,bach}
for recent reviews. Collagen ensures the mechanical stability and
strength of these tissues, but its biological role is certainly larger
because it also participates in biochemical and immunological protection
of the organism. Connective tissues are macroscopic hierarchical
structures that consist of several levels: a collagen tendon is made of
several fascicles held together. The fascicle consists of fibers, which
are made (going to the bottom of the hierarchy) from fibrils, micro-fibrils and collagen
triple-helices (which can be of several types, e.g., type I)
\cite{fratzl}. Each triple-helix consists of three poly-peptide chains
wound around each other. Fibrils and microfibrils are stabilized by
several different factors: a small number of covalent links between
terminal points of collagen triple-helices, carbonyl-water
hydrogen-bonds, hydrophobic and van der Waals interaction between
triple-helices \cite{fratzl,bach}. 

{\it Materials and Methods}. Collagen fibril samples with diameter from
1$\mu m$ and the length $1-0.1$ mm were extracted from Achilles tendons
of young rats by means of shaking and pulling using micro-tweezers in
$96 \%$ of ethyl alcohol at temperature 5 C. We call our sample fibril,
because its diameter is closer to the accepted diameter of the fibril
than to that of the fiber. Indeed, the fibril diameter established via
electron microscopy varies between 40 nm and 0.5$\mu m$ \cite{fratzl}.
For real fibrils this value is underestimated as the electron
microscopy demands drying of the collagen samples. 

The sample under investigation was enclosed in the experimental chamber
and placed in a temperature-controlled cabinet with the temperature
maintained at $25 ^{\circ}$C. The sample was allowed to equilibrate at a
given humidity for several hours. The relative humidities from $97$ to
$32\%$ in the chamber were achieved by means of $CaCl_{2}$ solutions of
different concentrations, while the relative humidities of $15\%$ and
$10\%$ were obtained via saturated solutions of $ZnCl_{2}$ and $LiCl$,
respectively. 

The Young's modulus $E$ and the logarithmic damping decrement $\theta$
were measured via electrically excited transverse resonance vibrations
of the sample (fibril cylinder), which is cantilevered from one edge
(another edge is free) \cite{morozov_gevorkian}. $E$ characterizes the
elasticity and defined as the ratio of pressure over strain
\cite{ferry,frenkel}; $\theta$ is determined from the amplitude of 
the sample oscillations and is defined as the ratio of the energy dissipated
during the externally forced oscillations to the energy (reversibly)
stored in the sample \cite{ferry,frenkel}. A large $\theta$
is typical for viscous liquids, while a small $\theta$ characterizes
elastic solids.

\begin{figure}
\includegraphics[width=8.5cm]{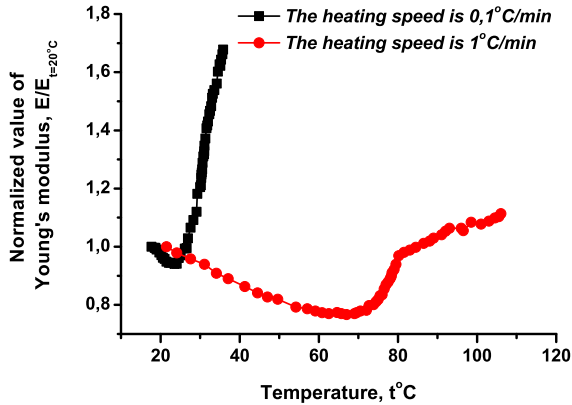}
\caption{The normalized Young's modulus $E$ of the collagen fibril versus temperature
for different heating speeds. The heating starts from 19 C and relative humidity $A=93\%$.}
\label{f3}
\end{figure}

\begin{figure}
\vspace{0.2cm}
\includegraphics[width=8.5cm]{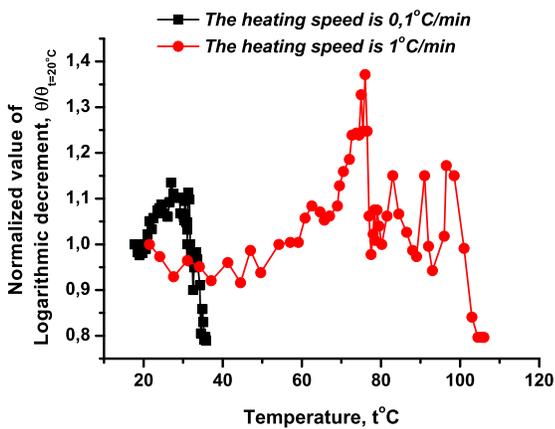}
\caption{The normalized logarithmic damping decrement $\theta$ of the collagen fibril versus temperature
for different heating speeds. The heating starts from 19 C and $A=93\%$.}
\label{f4}
\end{figure}

For measuring $E$ and the amplitude-frequency characteristics of
oscillations (employed for obtaining $\theta$), it is necessary to change
smoothly the frequency $f$ of the induced oscillations and determine the
resonance frequency, which corresponds to the maximal oscillation
amplitude of the sample's free end.  The Young's modulus of sample's
main axis is calculated via \cite{Lan}
\begin{equation}
\label{1}
E=3.19\cdot \omega^{2}\cdot L^{4}\cdot\rho\cdot P/I_{\rm min}, 
\end{equation}
where $\omega$ is the resonance frequency, $L$ is the sample length, $P$
is the cross-section area, $\rho$ is the density, and $I_{\rm min}$ is
the main inertia moment of that section, which corresponds to the
deformation plane with the minimal stiffness.  For the round
cross-section of our samples $I_{\rm min}={\pi\cdot D^{4}}/{64}$
\cite{Lan} and $P=\pi\cdot D^2/4$, where $D$ is the sample diameter
(measured with precision $0.02\,\mu$m).  Thus the Young's modulus is
obtained from (\ref{1}), where $L$, $\rho$, $P$ and $I_{\rm min}$ are
the known sample parameters and $\omega$ is measured on the experiment.
Studying samples with different length $L$, one can explore a range of
frequencies. 

Below we contrast the native features of collagen fibril with those of globular protein lysozyme (in solid state).
Tetragonal ($P4_3 2_1 2$) lysozyme crystals were grown following the
method of \cite{rost1} from a solution containing $3\%$ of lysozyme,
$5\%$ of $NaCl$, 0.2 M of natrium-acetate buffer, and ${\rm pH}=4.7$.
Crystals were fixated for 10 days via diffusion through the gas phase of
glutaraldehyde.  For another 10 days the crystalls were fixated
directly into $5\%$ glutaraldehyde solution with the same buffer.
Fixated crystalls have parameters of the elementary units ($a=b=79.4$ A
and $c=37.6$ A) close to native ones: $a=b=79.1$ A and $c=37.9$ A
\cite{rost2}.

{\it Mechanical glassy state.} Glass is a meta-stable state of matter,
which is caught in the process of (very) slow relaxation to equilibrium
\cite{ferry,frenkel}. Features of this state are sensitive to
observation times (frequencies). It is typically displayed whenever the
system does not have enough time to follow the external changes, e.g.,
when cooling with a small, but finite rate, or forcing the system with a
finite frequency. These two types of glassines are known, respectively,
as structural and mechanical \cite{ferry,frenkel}. 

During the mechanical glass transition the system is forced by an
external field at a frequency $\omega$ (with us this is the resonance
frequency of the sample vibration). When changing $\omega$ at a fixed
temperature $T$ (this is achieved by changing the sample length), 
the damping decrement $\theta(\omega)$|which 
characterizes internal friction and thus energy dissipation|displays a maximum
at some frequency $\omega_{\rm g}$ \cite{ferry,frenkel}.  Approximately
at the same frequency range $\omega\approx\omega_{\rm g}$, the Young's
modulus $E(\omega)$ (or some other elastic modulus) displays a crossover
between a relatively large high-frequency value $E_+$ to a relatively
small, low-frequency value $E_-$ \cite{ferry,frenkel}.  At the glass
transition frequency $\omega_{\rm g}$ (one of) the relaxation time(s) of
the systems becomes of order of $1/\omega_{\rm g}$ \cite{ferry,frenkel}.
In structural glass-formers the maximum of internal friction corresponds
to a large viscosity, while the two values $E_+$ and $E_-$ of $E$ refer
respectively to solid-like and liquid-like elaticity features.  The same
behavior of $E$ and $\theta$ is seen for a fixed frequency $\omega$
and the temperature changing at a small, but finite speed. The glass
transition now refers to the relaxation time being of the same
order as the inverse dimensionless temperature speed \cite{ferry,frenkel}. 

{\it Results and discussion.} Fig.~\ref{f1} displays the damping
decrement $\theta$ of the collagen fibril versus frequency $\omega$ and
at varying humidity and room temperature 25 C. Recall that this is the
resonance frequency of the sample at which we measure the Young's
modulus; we remind as well that the frequency varies with the sample
length. It is seen from Fig.~\ref{f1} that $\theta$ has a well-displayed
maximum at frequencies lower than $2$ kHz and that the behavior of
$\theta$ does not depend on the humidity. The plateau behavior of the
Young's modulus $E$ at the same temperature 25 C is seen on
Fig.~\ref{f2}. The plateau is most clealry visible at lowest studied
humidity, though it is identifiable for the highest humidity as well;
see Fig.~\ref{f2}. 

The results presented on Figs.~\ref{f1} and \ref{f2} imply that the room
temperature collagen fibril displays glassy features at room
temperature. The transition frequency $\omega_{\rm g}$ is smaller than
$2$ kHz at $T=25$ C. Changing the humidity does not alter $\omega_{\rm
g}$, in contrast to the low-temperature glassy features of globular
proteins, where the hydration level is known to be essential
\cite{review_1,review_2}. For the collagen fibril only the upper value
$E_+$ of the Young's modulus decreases (the fibril becomes more
flexible) for a higher humidity. To make the comparison with globular
proteins even more visible we studied visco-elastic features of
crystalline lysozyme at room temperature (each site of the tetragonal
crystal contains one globular lysozyme protein). Results presented in
Figs. \ref{f5}, \ref{f6} show no indications of glassy behavior.  It is
also seen that for the native lysozyme the dependence on the humidity is
essential. 

It is expected that upon heating the fibril will go out of the glassy
state. For synthetic polymers this phenomenon is known as the softening
transition \cite{ferry,frenkel}; this transition is accompanied by the
same effects as the glass transition: a peak of the internal friction and
a relatively abrupt change of the Young's modulus. Figs.~\ref{f3},
\ref{f4} show that when heating at speed 1 C/min the collagen fibril
undergoes softening transition at temperatures around $T_{\rm
soft}=70$ C, where $\theta$ has a sharp maximum. We note however that the
behavior of the Young's modulus $E$ is more intricate \cite{prl} than
during the glass transitions in synthetic polymers, e.g., rosin
\cite{ferry,frenkel}. It is seen that for $T<T_{\rm soft}$, $E$
decreases upon increasing $T$; this is a typical scenario for the
softening transition. However, for $T_{\rm soft}\lesssim T$ the Young's
modulus starts to increase with temperature. This effect was first found
in Ref.~\cite{prl}; it was argued to be related to formation of
inter-molecular bonds between partially denaturated fibril constituents.
Note that the denaturation temperature of collagen tendon, as measured
calorimetrically, is also located around $60-70$ C when heating under
speed larger than 1 C/min \cite{haly,miles,tiktopulo,tsereteli}.
Thus, around $70$ C there are three diferent processes taking place
almost simultaneously: partial denaturation of fibril, formation of
inter-molecular bonds which increases the Young's modulus for higher
temperatures, and the softening transition. Altogether, the collagen
physics around 70 C is complicated.  Now upon decreasing the heating
speed the softening transition should decrease, because for slower
heating a lower temperature suffice for displaying the relevant motion
within the observation times \cite{frenkel}. This is what we saw in our
experiments: for the heating speed 0.1 C/min we get $T_{\rm soft}\approx
25$ C; see Figs.~\ref{f3}, \ref{f4}.  This is the physiological
temperature regime for the native collagen. 
It is to be stressed that the glassy
features|including frequency-dependences and the softening transition|
are absent for the heat-denaturated collagen fibril, which is prepared
by keeping the native sample at $120$ C for several hours. Thus the
ordered character of the fibril is essential for displaying the glassy
features. We conjecture that these features are built in during the
fibrilogenesis; they can be similar to the {\it orientational glassines} 
phenomenon known for synthetic polymers \cite{frenkel}.

In this context it is interesting to note that the calorimetric studies
carried out in Refs.~\cite{miles,tsereteli} did indicate on the presence
of certain glassy feature in the heat-denaturated collagen tendon.  It was
also seen that these features are very sensitive to hydration changes,
and that they tend to disappear after annealing the heat-denaturated
collagen tendons \cite{tsereteli}.  There are here two essential
differences with respect to our results. First, we found glassy features
in the native (and not heat-denaturated) state of collagen fibril.
Second, the hydration level is not essential for these features, e.g.,
Figs.~\ref{f1}, \ref{f2} shows that changing the humidity level is not
essential for the characteristic glassy frequency $\omega_{\rm g}$. 

We close by repeating the main message of the present work: type I
collagen fibril is glassy at its physiological temperatures.  The
phenomenon is not sensitive to hydration changes, is specific to the
native state of the fibril, and constitutes the first example of
biopolymer glassiness at physiological temperatures. It may be of relevance
for understanding the genesis of collagen-based structures and for controlling 
their aging.

This work was supported by Volkswagenstiftung, ANSEF,
SCS of Armenia (grant 08-0166), and grants 96-2911-M 001-003-MY3 \& AS-95-TP-A07 of NSC-Taiwan.


\end{document}